\documentclass[12pt]{iopart}

\newcommand{\al}{\alpha}
\newcommand{\be}{\beta}
\newcommand{\ga}{\gamma}
\newcommand{\m}{\mu}
\newcommand{\n}{\nu}
\newcommand{\rh}{\rho}
\newcommand{\si}{\sigma}
\newcommand{\Si}{\Sigma}
\newcommand{\ta}{\tau}
\newcommand{\om}{\omega}
\newcommand{\mc}{\mathcal}

\newcommand{\pie}{\mbox{\Large$\pi$}}
\newcommand{\ce}{\mc{E}}
\newcommand{\cs}{\mc{S}}
\newcommand{\cl}{\mc{L}}
\newcommand{\ck}{\mc{K}}

\newcommand{\ca}{\mc{A}}
\newcommand{\p}{\perp}
\newcommand{\na}{\nabla}
\newcommand{\Rtwo}{^{^{\mbox{\tiny{(2)}}}}\hspace{-0.15cm}R}
\newcommand{\Rt}{^{^{\mbox{\tiny{(3)}}}}\hspace{-0.15cm}R}
\newcommand{\Qt}{^{^{\mbox{\tiny{(3)}}}}\hspace{-0.15cm}Q}
\newcommand{\nd}{\dot{n}}

\usepackage{amsbsy,amssymb}

\begin{document}

\title[Electromagnetic perturbations of non-vacuum LRS class II spacetimes]{Electromagnetic perturbations of non-vacuum locally rotationally symmetric class II spacetimes}

\author{R. B. Burston and A. W. C. Lun}

\address{School of Mathematical Sciences, Monash University, Australia 3800}
\eads{\mailto{Raymond.Burston@gmail.com}, \mailto{Anthony.Lun@sci.monash.edu.au}}
\begin{abstract}

We present a method that yields three decoupled covariant equations for three complex scalars, which completely govern electromagnetic perturbations of non-vacuum, locally rotationally symmetric class II spacetimes. One of these equations is equivalent to the previously established generalized Regge-Wheeler equation for electromagnetic fields. The remaining two equations are a direct generalization of the Bardeen-Press equations. The approach undertaken makes use of the well established 3+1 (and 2+1+1) formalism, and therefore, it is an ideal setting for specifying interpretable energy-momentum on an initial spacelike three-slice as the perturbation sources to the resultant electromagnetic radiation. 

\end{abstract}

\pacs{04.25.Nx, 04.20.-q, 04.40.-b, 03.50.De, 04.20.Cv}
%\submitto{\CQG}
\maketitle
\section{Introduction}

In this paper, we consider electromagnetic (EM) perturbations to non-vacuum background spacetimes that exhibit local rotational symmetry of class II \cite{Betschart, Ellis1967,Elst1996,Stewart}.

The {\it 3+1 slicing} formalism\cite{ADM} is employed as it has the desirable property of retaining a transparent and physical interpretation of the energy-momentum tensor. This corresponds to the ``zero rotation case" for the {\it 1+3 threading} formalism expounded by \cite{Ellis1973}. The four-dimensional spacetime is foliated into a family of spacelike three-slices, and subsequently, all tensors are decomposed irreducibly into spatial parts residing within the three-slices, and parts which are orthogonal.  As the assumption that the background is locally rotationally symmetric (LRS) class II is imposed, there is additional structure hidden within the equations, which is difficult to exploit in 3+1 form. To resolve this, a further splitting of the 3+1 quantities is desirable.  Every three-slice is therefore foliated by a family of two-slices, and all spatial quantities are decomposed into ``angular" parts residing entirely on the two-slices and ``radial" parts orthogonal to the two-slices. Analogous to the difference between the 3+1 and 1+3 formalisms stated above, this will be referred to as a {\it 2+1+1 slicing} formalism, and this is a special case of the {\it 1+1+2 threading} formalism developed in \cite{Clarkson,Clarkson2004}.  Ultimately, the LRS class II equations reduce to a system of partial differential equations (PDEs) involving only covariant scalar quantities\cite{Betschart}.

First we use linear algebra techniques to show that Maxwell's equations decouple naturally by constructing complex quantities. The  2+1+1 splitting is then used to decoupling the ``radial" quantities from the ``angular" quantities. The ``radial" quantity satisfies the Regge-Wheeler (RW) equation\cite{Regge} generalized to LRS class II spacetimes, for which the real and imaginary equations correspond to those derived in \cite{Betschart}.  

This paper extends the work in \cite{Betschart} by performing an additional covariant splitting of the all ``angular" quantities residing on the two-slices. Two covariant, decoupled, angular equations are ultimately derived, and we shown these are a direct generalization of the Bardeen-Press (BP) equations \cite{Bardeen} toward non-vacuum LRS class II spacetimes.

\subsection{Mathematical Preliminaries}
Greek indices run from $0$ to $3$, lower case Latin from $1$ to $3$ and capital Latin from $2$ to $3$.  The standard symmetric ( ) and skew-symmetric [ ] notation is used, and angle brackets $<>$ represent the spatially projected onto a three-surface, symmetric, and trace-free part of a tensor.  We define the four-coordinates $x^\m := (t,r,x^A)$, where $t$ is the ``temporal" coordinate, $r$ is the ``radial" coordinate and $x^A$ are the ``angular" coordinates. Note that, temporal, radial and angular are used generically here.  Geometric units are used whereby $ 8\pi G =c=1$. The completely anti-symmetric Levi-Civita pseudo-tensor $(\epsilon_{\m\n\si\ta})$ is defined such that $\epsilon_{\al\be\ga\om} \epsilon^{\al\be\ga\om}=-4!$.  The Lie derivative along a vector field $X^\m$ is denoted $\cl_X$.  The complex number is denoted by $\rmi$. The covariant derivatives with respect to the full spacetime, three-surface and  two-surface are respectively $(\na_\m, D_i, d_A)$. The Riemann tensors are therefore $2\,\na_{[\m} \na_{\n]} V_\si = R_{\m\n\si\ta} V^\ta$, $2\,D_{[\m} D_{\n]} W_\si =\Rt_{\m\n\si\ta} W^\ta$ and  $2\,d_{[\m} d_{\n]} V_\si =\Rtwo_{\m\n\si\ta} V^\ta$ where $W_\si$ has been projected onto the three-surface, and $V_\si$ has been projected onto the two-surface.  The Bianchi identities for a two-surface imply that the two-Ricci tensor can be written in terms of the Gaussian curvature ($K_g$) according to $\Rtwo _{\m\n} = K_g \cs_{\m\n}$ where $\cs_{\m\n}$ represents the metric of the two-surface.

\section{The Background LRS Class II Spacetime}

The background spacetime is assumed to be LRS class II. This section first presents the non-trivial equations for the 3+1 formalism and  subsequently, the 2+1+1 system of equations is discussed.

\subsection{3+1 Formalism}

The comprehensive 3+1 formalism for general relativity consists of both the Einstein field equations and the Bianchi identities. The 3+1 formalism is well established throughout the literature; for example, see \cite{ADM, Bernstein,Burston,York1979} and the ``zero rotation cases" of the 1+3 formalism in \cite{Bel1958,Ehlers1993,Ellis1967}. Thus only the essential mathematical tools required to decompose equations into 3+1 form are presented here, followed by a summary of the equations. The construction of spatial three-surfaces requires a time-like normal defined,
\begin{eqnarray}
 n_\m = -\al \na_\m t\qquad\mbox{and}\qquad n^\al n_\al    : = -1,
\end{eqnarray}
where $\al > 0$ is the lapse function and the Cauchy time function has, without loss of generality, been set equal to $t$. The projection tensor is used to project any quantity onto the three-surfaces and is defined,
\begin{eqnarray}\label{projection_tensor}
\p_{\m\n}: = g_{\m\n} +n_\m n_\n.
\end{eqnarray}
The associated three-metric permitted for a spatial three-slice is $\p_{ij}$.  The extrinsic curvature of the three-surfaces is related to the covariant derivative of the normal, which is decomposed according to,
\begin{eqnarray}
 \na_\m n_\n = -A_{\m\n}-\frac13\,\p_{\m\n}\,K -n_\m \nd_\n.
\end{eqnarray}
Here, $A_{\m\n}$ and $K$ are the {\it trace-free} and {\it trace} parts of the extrinsic curvature, and $\nd_\m := n^\al \na_\al n_\m$ is the four-acceleration. The ten independent components of the energy-momentum tensor ($T_{\m\n}$) are defined
\begin{equation}
\eqalign{  \rho:= T_{\al\be}n^\al n^\be ,\qquad P:= \frac 13 \p^{\al\be} T_{\al\be}, \cr
 j_\m :=-{\p_\m}^\be T_{\be\al}n^\al , \qquad
\pie_{\m\n}:={\p_\m}^\al {\p_\n}^\be T_{\al\be}-\frac 13 \p_{\m\n}\p^{\al\be}T_{\al\be} .}\label{eq2} 
\end{equation}
Here $\rh$ is the mass-energy density, $P$ is the isotropic pressure, $j_\m$ is the spatial mass-energy flux and $\pie_{\m\n}$ represents the spatial, trace-free, anisotropic pressure. 

The ADM equations \cite{ADM} comprise of both conservation of energy-momentum, and the Einstein field equations,
\begin{eqnarray}
\na^\al T_{\m\al} =0\qquad\mbox{and}\qquad   R_{\m\n}-\frac 12 g_{\m\n} \, R+\Lambda\, g_{\m\n}= T_{\m\n},
\end{eqnarray}
where $\Lambda$ is the cosmological constant.
 These can be decomposed into their 3+1 form and are respectively, the continuity and Euler equations,
\begin{eqnarray}
(\mc{L}_n-K)\, \rh +(D^k+ 2\,\nd^k)j_k -\pie_{km} A^{km} -P\, K=0,\label{continuity}\\
(\mc{L}_n -K) j_i + (D^k+\nd^k)\pie_{ik} +(D_i+\nd_i)\,P+\rh\,\nd_i = 0.
\end{eqnarray}
The Einstein field equations consist of the Hamiltonian and momentum constraints, and evolution equations for the extrinsic curvature,
\begin{eqnarray}
\rh=\frac 12 \Rt -\frac 12 A_{km} A^{km} +\frac 13 \,K^2-\Lambda,& \\
 j_i = D^k(A_{ik}-\frac 23\p_{ik}K),\\
 P =\frac 23 (\cl_n-\frac 13 \,K)  K -\frac 12 A_{km} A^{km}  +\frac 23\,(D^k+\nd^k) \nd_k-\frac 16 \Rt+ \Lambda ,\\
\pie_{ij}=   -(\cl_n-\frac 13 K) A_{<ij>}  -(D_{<i}+\nd_{<i})\nd_{j>}  +\Qt_{ij} \label{eqnforpiihj}.
\end{eqnarray}
Here, $^{(3)}Q_{ij}$  is the {\it trace-free} part of the three-Ricci tensor on a three-slice and $\Rt$ is the corresponding three-Ricci scalar.  The evolution equation for the three-metric is given by
\begin{eqnarray}
\mc{L}_n \p_{ij} =-2 \,A_{ij} -\frac 23 \p_{ij} K\label{evolutionforperp},
\end{eqnarray}
and additional equations of state for the energy-momentum are essential to close this system in general.

The ADM equations, given by (\ref{continuity})-(\ref{evolutionforperp}),  can be supplemented with the 3+1 gravito-electromagnetic (GEM) formalism \cite{Bel1958,Ellis1973,Maartens1998}. The Weyl tensor ($C_{\m\n\si\ta}$)  is split into two GEM tensors  defined according to
\begin{eqnarray}
E_{\m\n}:= C_{\al\m\be\n}n^\al n^\be\qquad\mbox{and}\qquad
B_{\m\n}:=\frac 12 \epsilon_{\m\be\ga} {C^{\be\ga}}_{\om\n}  n^\om \label{defintionforeandb},
\end{eqnarray}
where the completely anti-symmetric three-Levi-Civita tensor can be been defined to satisfy $\epsilon_{\m\n\si} := \epsilon_{\al\m\n\si} n^\al$.
The assumption of LRS class II spacetimes is sufficient to enforce the gravito-magnetic tensor ($B_{\m\n}$) to vanish everywhere\cite{Elst1996}. Therefore, using LRS class II symmetry, it follows that the non-trivial 3+1 GEM system\footnote{The 3+1 GEM system is the (once contracted) Bianchi identities decomposed into 3+1 form.} reduces to 
\begin{eqnarray}
D^k\left(E_{ki}+\frac 12 \pie_{ki}\right)= -\frac 12 \epsilon_{kim}{A^k}_n\left( \epsilon^{nmp}j_p\right) +\frac 1 3 D_i \rh + \frac 1 3  j_i \,K,\\
\left(\mc{L}_n-\frac 13 K\right)E_{<ij>}+\frac 12\epsilon_{km<i}\left(D^k+2\dot{n}^k\right)\left({\epsilon_{j>}}^{mn}j_n\right)\nonumber\\
+5 {A_{<i}}^k\left(E_{j>k}+\frac{1}{10}\pie_{j>k}\right)+ \frac 12\left(\mc{L}_n+\frac 13 K\right)\pie_{<ij>}=\frac 12(P+\rh)A_{ij}.
\end{eqnarray}
Finally, there are some additional relationships which yield useful information regarding the coupling between the ADM and GEM systems\footnote{(\ref{evforeij}) and (\ref{eqnforBij}) are found by first expressing the definition for the gravito-electric tensor (given in (\ref{defintionforeandb})) in terms of the Riemann tensor, Ricci tensor and Ricci scalar. Subsequently, this is decomposed to give an evolution equation for the extrinsic curvature. Finally, linear combinations with (\ref{eqnforpiihj}) gives the required results.},
\begin{eqnarray}
  E_{ij}  -\frac 12 \pie_{ij}=\cl_n A_{<ij>} + A_{k<i} {A_{j>}}^k+(D_{<i}+\nd_{<i})\nd_{j>},\label{evforeij}\\
E_{ij} +\frac 12 \pie_{ij} =\Qt_{ij} - A_{k<i} {A_{j>}}^k +\frac 13 A_{ij} K \label{eqnforBij}.
\end{eqnarray} 
The assumption of an LRS class II spacetime will greatly simplify these equations as most of them will have many trivial components, and this can certainly be analyzed from a component point of view. However, the covariant 2+1+1 formalism will be responsible for revealing only non-trivial scalar equations and this leads to simplifications.

\subsection{2+1+1 Decomposition}

In LRS class II spacetimes, every three-slice can be further foliated by a family of two-slices.  A normal to these two-slices $(N^\m)$ is spatial (i.e. $N_\al n^\al=0$), and  is normalized according to,
\begin{eqnarray}
 \perp^{\al\be}\,N_\al N_\be := 1,
\end{eqnarray}
from which it follows immediately,
\begin{eqnarray}
N_\m = \frac 1{\sqrt{\perp^{11}}}\, D_\m r,
\end{eqnarray}
where $ {\sqrt{\perp^{11}}}>0$ and $r$ is the ``radial" coordinate\footnote{The general solution for this normal is $N_\m =f\, D_\m g(r)$ where $f=f(t,r,x^A)$ and  $f\, g^\prime(r) = 1/\sqrt{\p^{11}}$. This is a similar situation to the Cauchy time function; $g(r)$ must be a function of $r$ only and can, without loss of generality, be set to equal $r$.}. A projection tensor which projects quantities onto the two-surfaces is therefore
\begin{eqnarray}
\mc{S}_{\m\n}:= \p_{\m\n} - N_\m\, N_\n .
\end{eqnarray}
The extrinsic curvature of the two-surfaces is related to the covariant three-derivative of the spatial normal,
\begin{eqnarray}
D_\m N_\n = -\mc{A}_{\m\n} -\frac 12 \cs_{\m\n} \,\mc{K}+N_\m \hat{N}_\n,
\end{eqnarray}
where $\mc{A}_{\m\n}$ and $\ck$ are the {\it trace-free} and {\it trace } parts of the extrinsic curvature of the two-surface, and $\hat {N}_\m:=N^\al D_\al N_\m$.

The 3+1 equations can now be decomposed into 2+1+1 form. The trace-free extrinsic curvature of the three-slice, the gravito-electric tensor and the anisotropic stress are decomposed as,
\begin{eqnarray}
A_{\m\n}= \Si_{\m\n}-\frac12\, S_{\m\n} \Si +2\, \Si_{(\m} N_{\n)}+\Si\,N_\m N_\n, \\
E_{\m\n}= \ce_{\m\n}-\frac12\, S_{\m\n} \ce +2\, \ce_{(\m} N_{\n)}+\ce\,N_\m N_\n, \\
\pie_{\m\n}= \Pi_{\m\n}-\frac12\, S_{\m\n} \Pi +2\, \Pi_{(\m} N_{\n)}+\Pi\,N_\m N_\n, 
\end{eqnarray}
where $\Si:= A_{\al\be}N^{\al }N^\be$, $\Si_\m:= {\cs_\m}^\al A_{\al\be} N^\be$, $\Si_{\m\n}={\cs_\m}^\al {\cs_\n}^\be A_{\al\be}+\frac 12 S_{\m\n}\, \Si$, and analogous definitions follow for $E_{\m\n}$ and $\pie_{\m\n}$.  It is important to note that now $\Si_{\m\n}$, $\ce_{\m\n}$ and $\Pi_{\m\n}$ are trace-free with respect to the two-slice; for example, $S^{\al\be} \Si_{\al\be}=0$.

The general 2+1+1 equations are not presented here, as only the LRS class II equations are required. LRS class II implies that the only non-vanishing terms are scalar quantities\cite{Betschart}, i.e.
\begin{eqnarray}
\mbox{LRS class II:}\,\, \{ \Si,K, \ca,\ce,\ck ,\Lambda,\rho, P,\mc{J},\Pi\},
\end{eqnarray}
where $\ca :=   \dot{n}_\al N^\al $ and $\mc{J} :=   j_\al N^\al$.  A more comprehensive 1+1+2 decomposition can be found in \cite{Betschart,Clarkson}.  Following the work of \cite{Betschart}, it is instructive to work with the Gaussian curvature of the two-surface,  
\begin{eqnarray}
K_g=-\ce+\frac 14 \ck^2-(\frac 13 K-\frac 12 \Sigma)^2 +\frac 13(\rho+\Lambda)-\frac 12 \Pi.
\end{eqnarray}
Therefore, the Ricci curvature for the three-surface can be decomposed according to
\begin{eqnarray}
\Rt =  2\,\left(\cl_N-\frac 34 \ck\right)\ck+2\,K_g,\\
\Qt_{\m\n} =-\frac 16\cs_{\m\n}(\cl_N \ck-2\,K_g)+\frac 13 N_\m N_\n (\cl_N\ck-2\, K_g).
\end{eqnarray}
In summary, the non-trivial LRS class II equations after irreducible decomposition are the continuity and Euler equations
\begin{eqnarray}
(\cl_n-\Si-\frac 43 \, K)\mc{J}+(\cl_N +\ca)P +(\cl_N +\ca-\frac 32\,\ck)\Pi +\rho \,\ca =0,\\
(\cl_n -K)\rh +(\cl_N +2\,\ca-\ck)\mc{J} - K\,P-\frac 32\, \Si\, \Pi =0.
\end{eqnarray}
The Hamiltonian and momentum constraints, the Raychaudhuri equation\cite{Raychaudhuri1955} and an evolution equation for  $\Si$ are respectively,
\begin{eqnarray}
\left(\cl_N-\frac 34 \ck\right)\ck+K_g+\frac 13 K^2-\frac 34 \,\Si^2= \rho+\Lambda, \\
\left(\cl_N-\frac 32 \ck\right)\Si-\frac 23\, \cl_N K =\mc{J},\\
(\cl_n-\frac 13 K)K+(\cl_N+\ca-\ck)\ca-\frac 32 \Si^2= \frac 12 (\rho+3P-2\Lambda),\\
-(\cl_n+\frac 32 \Si)(\Si-\frac 23 K)+(\cl_N -\ca-\frac 12 \ck)\ck= \Pi+\rho+P
\end{eqnarray}
The constraint and evolution equations governing the gravito-electric tensor are
\begin{eqnarray}
(\cl_N-\frac 32\ck)(\ce+\frac 12 \Pi)-\frac1 3 \cl_N\rho=-\frac 12(\Si -\frac 23 K) \mc{J} ,\\
\fl(\cl_n+\frac 32 \Si-K)\ce -\frac 13\cl_n\rho+\frac 12 (\cl_n+\frac 12 \Si-\frac 13 K)\Pi=-\frac 12 \ck\mc{J}+\frac 12(\rho+P)(\Si-\frac 23 K).
\end{eqnarray}
The equations giving information regarding the coupling of the ADM and GEM systems reduce to
\begin{eqnarray}
\ce-\frac 12\Pi=(\cl_n -\frac 13 K+\frac 12 \Si)(\Si-\frac 23 K)+\ca\ck+\frac 13(\rho+3\,P-2\,\Lambda),\\
 \ce+\frac 12 \Pi=(\cl_N -\frac 12 \ck)\ck-(\frac 13 K+\Si)(\Si-\frac 23 K)-\frac 23(\rho+\Lambda).
\end{eqnarray}
Finally, an additional evolution equation for $\ck$ is\footnote{This may be derived by decomposing $n^\m \cs^{\n\si} (2 \nabla_{[\m} \nabla_{\n]} N_\si - R_{\m\n\si\ta} N^\ta)=0$ as indicated in \cite{Clarkson}.},
\begin{eqnarray}
\left(\cl_n-\frac 13 K+\frac 12 \Si\right)\ck+\left(\Si-\frac 23 K\right)\ca =- \mc{J}.
\end{eqnarray}
These are the 2+1+1 equations governing the dynamics of the background spacetime\footnote{These equations correspond to those presented in \cite{Betschart,Clarkson} where the transformation from {\it our} notation to {\it theirs} is $\{ \Si,K,\ck,\rho, \mc{J},n_\m,N_\m\}\rightarrow\{- \Si,-K,-\phi ,\mu,\mc{Q},u_\m,n_\m\}$.}. They need to be prescribed in order to evaluate the perturbations.

\subsection{Commutators}
The decoupling of Maxwell's equations relies heavily on the commutation relationships between the Lie derivatives along both $n^\m$ and $N^\m$, and the covariant two-derivative on the two-surface. For any scalar function $f$, 
\begin{eqnarray}
 (\cl_N+\ca)\cl_n f -\bigl(\cl_n-\Si-\frac 13 \, K\bigr) \cl_N f   = 0,\label{com40}\\
d_\m\cl_n f  -\cl_n d_{\bar\m} f    =0   \label{com41},\\
d_\m \cl_N f- \cl_N d_{\bar\m} f =0,\label{com42}\\
d_{[\m} d_{\n]}  f =0.
\end{eqnarray}
Here, the ``bar" is used to denote an index which is projected onto the two-surface, i.e. $V_{\bar\m}:= {\cs_\m}^\al V_\al$. Similarly, for any two-tensor (i.e. $N^\al \Phi_\al=n^\al \Phi_\al =0$),
\begin{eqnarray}
[\cl_N+\ca+\frac 12 \ck][(\cl_n-\frac 12 \Si+\frac 13 K)\Phi_{\bar\m}]   = [\cl_n-\frac 32 \,\Si][(\cl_N+\frac 12 \ck)\Phi_{\bar\m}]   \label{com44},\\
   d_\m\cl_n \Phi_{\bar\n} - \cl_n d_{\bar\m} \Phi_{\bar\n}=0\label{com45},\\
d_\m \cl_N \Phi_{\bar\n}-\cl_N d_{\bar\m} \Phi_{\bar\n}=0\label{com46}.
\end{eqnarray}

\section{Maxwell's Equations}

There are two primary ideas discussed in this section. The first discusses Maxwell's field equations from a general point of view and illustrates the first stage of decoupling. The latter then focuses on the perturbed quantities and deriving the generalized RW and BP equations.

\subsection{Maxwell's Equations in General}\label{meigensdd}

The covariant 3+1 formalism for EM fields has been studied previously\cite{Burston,Ellis1973,Tsagas1997,Tsagas2005}. The spatial EM field intensities ($E_\m$ and $B_\m$)  are expressed covariantly in terms of the anti-symmetric Faraday tensor ($F_{\m\n}$),
\begin{eqnarray}
E_\m := F_{\m\al}n^\al \qquad\mbox{and}\qquad B_\m :=\frac 12 \epsilon_{\m\al\be} F^{\al\be}. 
\end{eqnarray}
It then follows that the EM tensor is decomposed into 3+1 form according to
\begin{eqnarray}
F_{\m\n}=\epsilon_{\m\n\al}B^\al - 2\,E_{[\m} n_{\n]} .
\end{eqnarray}
Maxwell's equations are expressed conveniently in terms of $F_{\m\n}$ through 
\begin{eqnarray}
\na_{[\si}F_{\m\n]}=0 \qquad\mbox{and} \qquad\na^\al F_{\m\al}&=\frac 12 J_\m. 
\end{eqnarray}
Here $J_\m$ represents the charge-current four-vector which is decomposed as
\begin{eqnarray}
 J_\m =\p J_\m -(n^\al J_\al) n_\m = i_\m +\epsilon n_\m,
\end{eqnarray}
where $\p J_\m :=i_\m$ is the three-current density and $n^\al J_\al:=-\epsilon$ is the charge density. Maxwell's equations can now be expressed in 3+1 form,
\begin{eqnarray}
  D^k B_k  =   0, \label{divbezseo}\\
                D^k E_k  =  \frac 12 \epsilon , \\
(\mc{L}_n-\frac 13 K) B_i + {\epsilon_i}^{km}(D_k+\nd_k) E_m  +2\, {A_i}^k \,B_k = 0 ,\\
(\mc{L}_n-\frac 13 K)E_i   - {\epsilon_i}^{km}(D_k +\nd_k)  B_m   + 2\, {A_i}^k \, E_k= -\frac 12\, i_i   \label{liee}.
\end{eqnarray}
This is a coupled system of two constraint plus six evolution equations for the six EM field components.  One traditional approach to decouple the equations is to attempt to construct a second-order differential equation for each of $E_i$ and $B_i$ and it is shown in \cite{Betschart} that some of the 2+1+1 EM perturbations do satisfy decoupled equations.  In this paper, an alternative method is described which proves to be successful in decoupling the entire system. We seek to decouple this system naturally by using the linear algebra techniques specified in \ref{LADe}. There exists a complex conjugate pair of combinations of $E_i$ and $B_i$ which will naturally decouple the system, and a new complex tensor is defined accordingly,
\begin{eqnarray}
\Psi_\m :=  E_\m + \rmi\, B_\m.
\end{eqnarray}
The system \eref{divbezseo}-\eref{liee} is then given by,
\begin{eqnarray}
                D^k \Psi_k  =  \frac 12\, \epsilon ,\label{divpsi}\\
(\mc{L}_n-\frac 13 K)\Psi_i   + 2\,{A_i}^k\, \Psi_k +\rmi\, {\epsilon_i}^{km}(D_k +\nd_k )  \Psi_m   =-\frac12\, i_i \label{liepsi}   .
\end{eqnarray}
It is simple to show that the system for the complex conjugate ($\bar\Psi_i$) can be found by taking the complex conjugate of (\ref{divpsi}) and (\ref{liepsi}).

The decoupled system for the EM fields can be determined entirely from either $\Psi_i$ or $\bar\Psi_i$. Each system involves four complex equations governing three complex quantities.  The complex nature of these quantities is arising due to the invariant structure of the source less equations under the simultaneous transformation $(E_\m\rightarrow B_\m, B_\m \rightarrow-E_\m)$.  The use of a complex combination of the EM fields is well established throughout the literature and dates as far back as  \cite{Waelsch1913}.  The purpose of this first stage of decoupling is to emphasise that the use of complex quantities is not a random choice, but instead a natural construction arising due to the inherent structure of the equations.

The first stage decoupling has been completed in that now we have a system involving one tensorial quantity. The next challenge is to decouple the individual components of $\Psi_i$.  The system for $\Psi_i$ can be decomposed according to the $2+1+1$ decomposition,  
\begin{eqnarray}
(\mc{L}_N -\ck)\Phi+ d^A\Phi_A=\frac 12 \epsilon,\label{divons}\\
(\mc{L}_n -\frac 23\, K+\Si)\Phi+\rmi\,\epsilon^{AB}\,d_A\Phi_B=-\frac 12 \mc{I},\label{divsec}\\
\left(\cl_n-\frac 13 K-\Si\right)\Phi_{\bar A}-\rmi \,{\epsilon_A}^B\,(\cl_N+\mc{A})\Phi_B+\rmi \,{\epsilon_A}^B\,d_B\Phi=-\frac12 \mc{I}_A,\label{thirdone}
\end{eqnarray}
where $\Psi_\m     = \Phi_\m + \Phi\, N_\m$, $\Phi := N^\al \Psi_\al$, $\Phi_\m := {\cs_\m}^\al \Psi_\al$ and the three-current has been decomposed as $i_\m = \mc{I} N_\m + \mc{I}_\m$.  The parity reversed form of (\ref{thirdone}) is derived by contracting it with completely anti-symmetric the two-Levi-Civta tensor (which is defined such that $\epsilon_{\m\n} = \epsilon_{\m\n\al} N^\al$),
\begin{eqnarray}
(\cl_N+\mc{A})\Phi_A-\rmi\,{\epsilon_A}^B\left(\cl_n-\frac 13 K-\Si\right)\Phi_{ B}-\,d_A\Phi=\rmi\,\frac12 {\epsilon_A}^B\,\mc{I}_B.\label{paritythirdone}
\end{eqnarray}
Ultimately, it is desirable to find a decoupled equation for $\Phi$ such that the wave operator is present. Therefore, it is required to know how the scalar wave operator appears in both 3+1 and 2+1+1 form\cite{Betschart} which are respectively,
\begin{eqnarray}
 \square\phi    &:=& \na^\al \na_\al \phi \nonumber  \\
      &= &\left( -\left(\mc{L}_n -K\right) \mc{L}_n + \left(D^k+\nd^k\right) D_k\right) \phi \nonumber\\
    &=& \left( -\left(\mc{L}_n -K\right) \mc{L}_n +(\cl_N-\ck+\ca)\cl_N+ d^\al d_\al\right) \phi \label{tpopowaveoperator}
\end{eqnarray}
where $\phi$ is any scalar function.

In the fully non-linear case, Maxwell's equations are heavily coupled to the ADM and GEM equations through extrinsic curvature and acceleration terms appearing in the equations, and through the energy-momentum tensor. We now discuss how they simplify under a first-order perturbation.

\subsection{EM Perturbations to Non-Vacuum LRS Class II Spacetime}

The EM perturbations to an LRS class II spacetime consists of both EM fields and charged sources, i.e. $\Phi$, $\Phi_A$, $\epsilon$, $\mc{I}$ and $\mc{I}_A$ become first-order quantities. This paper assumes that these first-order EM fields and charged sources are gauge invariant\footnote{They are gauge invariant under infinitesimal coordinate transformations.}. Therefore, due to the Stewart-Walker lemma\cite{Stewart1974}, they vanish on the background. This consequently places a restriction on the {\it background} energy-momentum ($\rho,P,\mc{J},\Pi$) that it also be non-charged.  

It is also important to consider how these first-order EM fields  and sources feedback to the geometry. For example, a typical energy-momentum tensor for these EM perturbations, which includes a simple interaction term, is \cite{Barut}
\begin{eqnarray}
T_{\m\n}= F_{\m\al} {F^\al}_\n -\frac 14 \,g_{\m\n}\,F_{\al\be} F^{\al\be} + g_{\m\n} \mc{A}_\al J^\al ,
\end{eqnarray}
where $\mc{A}_\m$ is a four-potential defined $F_{\m\n} :=2\, \na_{(\m} \mc{A}_{\n)}$. This clearly involves second-order terms at most and will therefore vanish when truncating at first-order.  This simplifies the calculations as there are no first-order gravitational effects that need to be accounted for, and thus the geometry remains fixed.

\subsection{Generalized Regge-Wheeler Equation}

The outline for the derivation of the decoupled covariant equation for $\Phi$ is given here. First take the Lie derivative with respect to $n^\m$ of (\ref{divsec}) and use \eref{com45} to interchange the Lie derivative with respect to $n^\m$ and two-derivative acting on $\Phi_A$,  then \eref{thirdone} is substituted for $\cl_n \Phi_A$. Then \eref{com46} is used to interchange the Lie-derivative with respect to $N^\m$ and the two-derivative when acting on $\Phi_A$. Finally, \eref{divons} and \eref{divsec} are substituted to eliminate any remaining $\Phi_A$ terms to give,
\begin{eqnarray}
\fl\left[(\cl_n+\Si-\frac 53\, K)\cl_n -\left(\cl_N +\ca-2\,\ck\right)\cl_N  -d^\al d_\al -2\,K_g +\rho-P-\Pi +2\,\Lambda\right]\Phi\nonumber\\
=\rmi\,\frac 12\, \epsilon^{\al\be} d_\al \mc{I}_\be-\frac 12\,\left[(\cl_N+\ca-\ck) \, \epsilon+(\cl_n -K)\,\mc{I}\right].
\end{eqnarray}

This is the Regge-Wheeler (RW) equation\cite{Regge} generalized to LRS class II spacetimes with a non-charged background fluid.  There are actually two independent equations here, one for each of the real and imaginary components of $\Phi$, which correspond to those developed in \cite{Betschart}\footnote{The corresponding equations presented by \cite{Betschart} have slightly different numerical factors residing in the potential. The results presented here have been checked by expressing the equation using both the Schwarzschild metric (for which $\ca,\ck,\ce\ne0$ and $\Si,K=0$) and the Painleve Gullstrand metric (for which $\ca=0$ and $\ck,\,\ce,\,\Si,\,K\ne0$) and the corresponding coordinate transformation yields consistent results. This cannot be achieved using \cite{Betschart}'s results. However, it is of primary importance that \cite{Betschart}'s decoupling is not affected.}.   

By inspection of the differential operator above with (\ref{tpopowaveoperator}), there are clearly additional derivative terms left over once the true scalar wave operator has been identified.   It is possible to rescale $\Phi$ to obtain a true wave equation with a potential and a source defined in terms of energy-momentum quantities on the spacelike three-slices. This procedure is also presented in \cite{Betschart} and the scaling is given by
\begin{eqnarray}
\fl\mc{M} := \Phi\, \exp{(\Omega)}\qquad\mbox{where}\qquad\cl_n\Omega := \frac 12\Si-\frac 13 \,K \qquad\mbox{and}\qquad \cl_N \Omega := -\frac 12\ck.
\end{eqnarray}
It is important to show that these equations for $\Omega$ are in fact integrable,  i.e. they must satisfy \eref{com40},
\begin{eqnarray}
 (\cl_N+\ca)\cl_n \Omega -\bigl(\cl_n-\Si-\frac 13 \, K\bigr) \cl_N \Omega   = 0.
\end{eqnarray}
which is achieved using the 2+1+1 field equations. Under this scaling, the  usual scalar wave operator  appears with additional potential terms arising,
\begin{eqnarray}
(\square -V)\mc{M} =S,
\end{eqnarray}
where
\begin{eqnarray}
\fl V:
=-2\,K_g -\frac 12\,(\frac 12\,\Sigma-\frac 43\, K)(\Sigma-\frac 23\,K) -\frac 12\, (\ca-\frac 32\,\ck)\ck+\rho-P-\Pi +2\,\Lambda,\\
S:=\frac 12 e^{\Omega} \Bigl[-\rmi\, \epsilon^{\al\be} d_\al \mc{I}_\be+(\cl_N+\ca- \frac 12\ck) \, \epsilon+(\cl_n -K)\,\mc{I}\Bigr].
\end{eqnarray}
 In the absence of charged sources ($S=0$), this wave equation may be separated out into a time/radial part and angular part. This is a because the potential is a function of $t$ and $r$ only. Such vacuum solutions can then be found using standard  techniques; for example, spherical harmonic solutions arise for the angular components. These vacuum solutions offer invaluable information regarding how the EM waves interact  with their background surroundings.  However, in the presence of charged sources, this gives the exciting prospect of identifying the EM signature from a particular source.

\subsection{The Generalized Bardeen-Press Equations}

We now extend the results presented in \cite{Betschart} to show that the natural construction of complex quantities also yields a decoupled equation for $\Phi_A$. The derivation is analogous to  how the RW equation was constructed. Take the Lie derivative with respect to $n^\m$ of \eref{thirdone} and use \eref{com41} to interchange the Lie derivative with respect to $n^\m$ and the two-derivative when acting on $\Phi$, and then substitute \eref{divsec} to eliminate $\cl_n \Phi$. Also use \eref{com44} to interchange the two Lie derivatives when acting on $\Phi_A$ and subsequently, substitute \eref{thirdone} to eliminate $\cl_n \Phi_A$.  Then use \eref{com42} to interchange the Lie derivative with respect to $N^\m$ and the two-derivative when acting on $\Phi$, and use \eref{divons} to eliminate $\cl_N \Phi$. Finally, use \eref{thirdone} and its parity relationship (\ref{paritythirdone}) to eliminate any remaining two-derivatives acting on $\Phi$, to show that
\begin{eqnarray}
\fl \left[(\cl_n-K)\cl_n-(\cl_N+\ca- \ck)\cl_N- d^B d_B)\right]\Phi_{\bar A} -\rmi \,{\epsilon_A}^B \,\left[(2\,\ca+\ck)\cl_n   +3 \, \Si\,\cl_N\right]\Phi_B \nonumber \\
+\left[K_g-(\cl_n+\Si-\frac 23 K)(\Si+\frac 13 K)-(\cl_N-\ck)\ca\right]\Phi_A \nonumber\\
  -\rmi\,{\epsilon_A}^B\Bigl[(\cl_n-K)(\ca-\frac 12 \ck) +(\cl_N+\ca-\ck)(\frac 12 \Si+\frac 23 K)\Bigr]\Phi_B\nonumber\\
   =\rmi\,\frac 12\,{\epsilon_A}^B \left[d_B \mc{I} -(\cl_N-\ck)\mc{I}_B \right] -\frac 12\left[(\cl_n+\Si-\frac 23 \, K) \mc{I}_{\bar A}- d_A \epsilon\right] \label{waveforphia}
\end{eqnarray}
This covariant complex equation has clearly decoupled from $\Phi$, and the next difficulty is to show how to decouple the two independent components of $\Phi_A$.

This equation can be expanded in component form, by specifying an {\it arbitrary LRS class II form of the metric}, and then expressed in terms of large, untidy,  matrices. The natural decoupling methodology used in Section \ref{meigensdd} and \ref{LADe} can be employed again, and this again demands that complex combinations be formed to naturally decouple the equations. 

However, {\it it is not necessary to specify the metric at all}. Instead, a natural covariant decomposition of all two-surface quantities can be achieved by first defining a complex-conjugate pair of vectors $(m^\m, {\bar{m}}^\m)$ which satisfies the following relationships:
\begin{eqnarray}
{\bar m}^\al m_\al =1,\qquad m^\al m_\al =0, \qquad {\bar m^\al }{\bar m_\al} =0,\qquad S^{\m\n} = 2\, m^{(\m} {\bar m}^{\n)}.\label{mbarprops}
\end{eqnarray}
These complex-conjugate vectors are orthogonal to both $n^\m$ and $N^\m$. Consequently, they can be raised and lowered using the two-metric.

The remaining equations on the two-surface can now be irreducibly decomposed.  Consider any two-tensors $V_\m$, $V_{\m\n}$ (i.e. a contraction of any of their indices with $n^\m$ or $N^\m$ will vanish), where $V_{\m\n}$ is symmetric and trace-free with respect to the two-surface ($\cs^{\al\be}V_{\al\be}=0$). By using the properties specified in \eref{mbarprops}, they are both irreducibly decomposed according to
\begin{eqnarray}
V_\m = (V_\al m^\al) \bar m_\m + ( V_\al \bar m^\al) m_\m,\\
V_{\m\n} = ( V_{\al\be} \bar m^\al \bar m^\be)  m_\m  m_\n+( V_{\al\be} m^\al m^\be) \bar m_\m \bar m_\n,
\end{eqnarray}
and the generalization to tensors of different type is evident.
 Therefore, $\Phi_A$ is irreducibly  decomposed as
\begin{eqnarray}
\Phi_A  =  \mc{M}_\oplus\, {\bar m_A} + \mc{M}_\otimes\,m_A,\label{angdeophiq}
\end{eqnarray}
where the scalars are defined $\mc{M}_\oplus:= m^A \Phi_A$ and $\mc{M}_\otimes := {\bar m^A} \Phi_A$. It is always possible to perform a rotation such that  $m^\m \rightarrow e^{-\rmi\,\varphi}m^\m$, which, by using \eref{mbarprops},  clearly leaves the two-metric invariant. Under this rotation both $\mc{M}_\oplus$ and $\mc{M}_\otimes$ are not invariant. However, their reconstruction back into $\Phi_A$ using \eref{angdeophiq} is invariant.

The two decoupled covariant equations arising naturally from the decomposition of \eref{waveforphia} are,
\begin{eqnarray}
\fl[\cl_n+2\ga -K-q\,(2\,\ca+\ck)]\cl_n \mc{M}_\oplus - \left[\cl_N+2\,\lambda+\ca-\ck+3\,q\,\Si\right]\cl_N \mc{M}_\oplus-d^\al d_\al \mc{M}_\oplus\nonumber\\
\fl-2\, \chi^\al d_\al \mc{M}_\oplus 
+\Bigl[(\cl_n-K +\ga)\ga-p\,(2\ca+\ck)-(\cl_N+\ca-\ck+\lambda)\lambda -3\,\omega\,\Si +K_g\nonumber\\
\qquad-q\,(\cl_n-K)(\ca-\frac 12 \ck) -q\,(\cl_N+\ca-\ck)(\frac 12 \Si+\frac 23 K)+\chi-d^\al\chi_\al\nonumber\\
\qquad-(\cl_n+\Si-\frac 23 K)(\Si+\frac 13 K)-(\cl_N-\ck)\ca\Bigr]\mc{M}_\oplus=\frac 12 S_\oplus\label{BPones},
\end{eqnarray}
\begin{eqnarray}
\fl[\cl_n+2\bar \ga -K+ q\,(2\ca+\ck)]\cl_n \mc{M}_\otimes - \left[\cl_N+2\,\bar\lambda+\ca-\ck-3\,q\,\Si\right]\cl_N \mc{M}_\otimes -d^\al d_\al \mc{M}_\otimes\nonumber\\
\fl+2\, \chi^\al d_\al \mc{M}_\otimes+\Bigl[(\cl_n-K+\bar \ga)\bar\ga-\bar p\,(2\ca+\ck)-(\cl_N+\ca-\ck+\bar\lambda)\bar \lambda-3\,\bar \omega\,\Si+K_g\nonumber\\
\qquad+q\,(\cl_n-K)(\ca-\frac 12 \ck) +q\,(\cl_N+\ca-\ck)(\frac 12 \Si+\frac 23 K)+\chi+ d^\al\chi_\al\nonumber\\
\qquad-(\cl_n+\Si-\frac 23 K)(\Si+\frac 13 K)-(\cl_N-\ck)\ca\Bigr]\mc{M}_\otimes=\frac 12 S_{\otimes}\label{BPtwos},
\end{eqnarray}
where the source has been decomposed as $S_\m:= S_\oplus \,\bar m_\m+ S_\otimes \,m_\m$.  The newly defined coefficients arising in these equations are related to various derivatives and combinations of $m_\m$ and $\bar m_\m$. They are given in \ref{ccrelations} for LRS class II, and it should be noted that in most cases they are related to other geometric LRS class II scalars. 

The covariant decoupled equations, \eref{BPones} and \eref{BPtwos}, are the direct generalization of the Bardeen-Press equations \cite{Bardeen,Chandra,Fernandes,Penrose} to non-vacuum LRS class  II spacetimes. In the absence of charged sources ($S_\oplus=S_\otimes=0$), the operators are separable which is due to the fact that the LRS class II scalars are functions of $t$ and $r$ only and the two-metric is expressed in terms of the complex-conjugate vectors. Such vacuum solutions may be obtained using standard techniques. In this case though, the angular solutions will be in terms of spin-weighted spherical harmonics\cite{Penrose}.  Furthermore, it is possible to specify energy-momentum on initial three-slices as the sources to the EM radiation.

\subsection{Example: Static Schwarzschild Background}

We now discuss the static Schwarzschild background to show that \eref{BPones} and \eref{BPtwos} are a generalisation of the BP equations. The static Schwarzschild line element is
\begin{eqnarray}
ds^2= -\left(1-\frac{2\,M}r\right)dt^2 +\frac 1 {\left(1-\frac {2\,M}r\right)} dr^2 + r^2\left(d\theta^2+\sin^2\theta d\phi^2\right),
\end{eqnarray}
and the 2+1+1 background equations reduce to 
\begin{eqnarray}
\fl(\cl_N+\ca-\ck)\ca=0,\qquad (\cl_N -\ca-\frac 12\ck)\ck=0\qquad\mbox{and} \qquad \ce=\ca\,\ck.
\end{eqnarray}
Therefore, the only non-zero components of the LRS class II scalars are,
\begin{eqnarray}
\fl\ca=\frac M{r^2}\left(1-\frac{2\, M}{r}\right)^{-\frac 12},\qquad \ck=-\frac 2 r \sqrt{1-\frac {2\, M} r}\qquad\mbox{and}\qquad\ce=-\frac {2\, M}{r^3}.
\end{eqnarray}
Furthermore,  the complex-conjugate vectors defining the two-metric is defined
\begin{eqnarray}
m^\m =\frac 1{\sqrt{2}\, r }\left(0,0,1,\rmi\, \mbox{cosec}\,\theta\right)\label{ccvector},
\end{eqnarray}
but this is not a unique choice.
Before proceeding, it is noted that the decoupled equations were checked by comparing them with those derived using the  Newman-Penrose formalism\cite{Chandra}. The null vectors used for this purpose were defined according to \cite{Chandra} and they are also not a unique choice. Therefore, to put the equations in the exact same form as those presented in \cite{Chandra},  one final scaling is required\footnote{In general the scaling is not necessary. The equations were checked by expanding everything in terms of coordinates using Maple 9.5 and  they correspond exactly to those derived using the  Newman-Penrose formalism\cite{Chandra}.},
\begin{eqnarray}
 \mc{M}_\oplus := f\,  \mc{B}_\oplus\qquad\mbox{and}\qquad  \mc{M}_\otimes := \frac 1f\,  \mc{B}_\otimes,
\end{eqnarray}
where $\cl_N \ln f = \ca$ and $\cl_n f=d_\m f =0$.
Finally, including the RW equation, we have three decoupled equations,
\begin{eqnarray}
\fl\left[\cl_n\cl_n -\left(\cl_N +\ca-2\,\ck\right)\cl_N  -d^\al d_\al -V_{\mbox{\tiny{(RW)}}}\right] \Phi=0
,\label{RW}\\
\fl \left[(\cl_n -2\,\ca-\ck)\cl_n - \left(\cl_N+3\,\ca-2\,\ck\right)\cl_N -d^\al d_\al-2\, \chi^\al d_\al  -V_\oplus \right] \mc{B}_\oplus=0,\label{BP1}\\
\fl \left[(\cl_n+ 2\ca+\ck)\cl_n  -\, \left(\cl_N\,-\,\,\ca-2\,\ck\right)\cl_N-d^\al d_\al +2\, \chi^\al d_\al -V_\otimes\right]\mc{B}_\otimes=0\label{BP2}.
\end{eqnarray}
Here the Lie derivatives reduce to $\cl_n =\frac 1\al \frac\partial{\partial t}$ and $\cl_N =  {\sqrt{\perp^{11}}} \frac\partial{\partial r}$, and the potentials are
\begin{eqnarray}
V_{\mbox{\tiny{(RW)}}} :=2\, K_g,\\
 V_\oplus := -\frac 12(\cl_N+ \ca-\ck)\ck -\chi=K_g-\chi,\\
 V_\otimes := -\frac 12(\cl_N-3\,\ca-\ck)\ck-\chi=K_g+2\,\ce-\chi.
\end{eqnarray}
The RW is given by  \eref{RW} for the quantity $\Phi$ which has zero spin-weight. Furthermore, \eref{BP1} and \eref{BP2} are precisely the BP equations \cite{Bardeen,Chandra,Fernandes,Penrose} for the spin-weighted quantities of $+1$ and $-1$ respectively.

\section{Discussion}

We have shown how to derive a fully decoupled set of three covariant equations, which completely determines the EM perturbations about a non-vacuum LRS class II spacetime. One of these is the generalized Regge-Wheeler equation, whereas the other two are the generalization of the Bardeen-Press equations. The setting uses the 2+1+1 formalism, and consequently, the perturbations are rich with physical significance. The equations can be expressed in terms of quantities such as the Gaussian and extrinsic curvature of two-dimensional surfaces thereby rendering the interpretation relatively simple. In vacuum perturbations, all the differential operators are separable and solutions can be found using separation of variables, and the usual spherical harmonics and spin-weighted spherical harmonics naturally arise. Furthermore, the sources to the equations are written in terms of charges and currents specified on initial spacelike three-slices.  This gives the exciting prospect of modelling EM radiation from a particular astrophysical source, rather than the more abstract vacuum pertubations.

\appendix

\section{Complex-Conjugate Identities}\label{ccrelations}

In LRS class II the non-trivial identities and definitions are:
\begin{eqnarray}
\eqalign{\ga := m^\al \cl_n \bar{m}_\al,\qquad\ga +\bar \ga  = \Si-\frac 23 K,\qquad\cl_n m^\al \cl_n \bar m_\al =-\ga^2,\\
\lambda := m^\al \cl_N \bar{m}_\al, \qquad   \lambda +\bar \lambda =   -\ck,\qquad\qquad \cl_N m^\al \cl_N \bar m_\al =-\lambda^2,\\
\chi_\m := m^\al d_\m \bar m_\al,\qquad\chi := (d^\al m^\be)(d_\al \bar m_\be),\qquad p:= \rmi\, m_\al \epsilon^{\al\be} \cl_n \bar m_\be,\\
q:= \rmi\, m_\al \bar m_\be \epsilon^{\al\be}=\pm1\qquad\mbox{and}\qquad \omega:= \rmi\, m_\al\epsilon^{\al\be} \cl_N \bar m_\be.}
\end{eqnarray}
In the static Schwarzschild spacetime, with the complex-conjugate vector defined by \eref{ccvector}, they reduce to
\begin{eqnarray}
\eqalign{ \ga =\bar \ga  =p=\bar p=d^\al \chi_\al=0,\qquad      \lambda =\bar \lambda =  \omega=\bar \omega = -\frac 12 \ck,\\
q=1,\qquad \chi_\m = [0,0,0,\rmi\,\cos\theta]\qquad\mbox{and}\qquad\chi=\frac{\cot^2\theta}{r^2}.}
\end{eqnarray}

\section{Linear Algebra: Decoupling}\label{LADe}

Consider the system given by 
\begin{eqnarray}
L_1 \,E+L_2\,B =0 \qquad\mbox{and}\qquad L_1\,B-L_2\,E=0,
\end{eqnarray}
where $L_1$ and $L_2$ represent differential operators and $E$ and $B$ are any scalar fields.  This system has the property that it is invariant under the simultaneous transformation of $E\rightarrow B$ and $B\rightarrow-\,E$. This system can be expressed in  a matrix form as
\begin{eqnarray}\label{matrix_one}
 \left(\begin{array}{c}
L_1\,E     \\
L_1\,B 
\end{array}\right) +  \left(\begin{array}{cc}
 0     &1    \\
   -1& 0 
\end{array}\right) \left(\begin{array}{c}
L_2\,E     \\
L_2\,B 
\end{array}\right)= \left(\begin{array}{c}
 0         \\
    0 
\end{array}\right),
\end{eqnarray}
where $M := \left(\begin{array}{cc}
 0     &1    \\
   -1& 0 
\end{array}\right) $
is the matrix responsible for coupling $E$ to $B$. $M$ can be written in terms of its eigenvalues, $\mbox{diag}({D})$, and corresponding eigenvectors, col($P$), according to $M = P \, D\,P^{-1}=\left(\begin{array}{cc}
 \rmi     &-\rmi    \\
     1 & 1 
\end{array}\right)\left(\begin{array}{cc}
- \rmi     &0   \\
   0& \rmi
\end{array}\right)\frac12\left(\begin{array}{cc}
-\rmi     &1    \\
   \rmi& 1 
\end{array}\right).
$
Therefore, since $D$ is diagonal, it is clear that by multiplying  (\ref{matrix_one}) by $-2\,\rmi\,P^{-1}$ results in the decoupled system,
\begin{eqnarray}
\fl  L_1(E+\rmi\,B)+\rmi\,\,L_2(E+\rmi\,B)=0\qquad\mbox{and}\qquad L_1(E-\rmi\,B)-\rmi\,\,L_2(E-\rmi\,B)=0.
\end{eqnarray}
Thus a complex-conjugate pair of equations arise. This result can be generalised to tensors of any type without loss of generality provided the invariance is satisfied. The matrix $M$ needs to be written in block form with blocks of zeros or the identity matrix to compenstate for the number of dimensions.

\end{document}